\documentclass[mathleft]{an}
\usepackage{graphicx}
\usepackage{times}
\begin{document}

\Pagespan{789}{}
\Yearpublication{2009}%
\Yearsubmission{2005}%
\Month{11}%
\Volume{999}%
\Issue{88}%

\title{The effect of clouds on the dynamical and chemical 
evolution of gas-rich dwarf galaxies}

\author{S. Recchi\inst{1}\fnmsep\thanks{simone.recchi@univie.ac.at}
\and  G. Hensler\inst{1}\fnmsep\thanks{gerhard.hensler@univie.ac.at}
}
\titlerunning{Clouds and evolution of gas-rich dwarf galaxies}
\authorrunning{S. Recchi \& G. Hensler}
\institute{Institute of Astronomy, University of Vienna, 
                 T\"urkenschanzstrasse 17, A-1180 Vienna, Austria}

\received{}
\accepted{}
\publonline{later}


\abstract{We study the effects of clouds on the dynamical
          and chemical evolution of gas-rich dwarf galaxies, in
          particular focusing on two model galaxies similar to IZw18
          and NGC1569.  We consider both scenarios, clouds put at the
          beginning of the simulation and continuously created
          infalling ones.  Due to dynamical processes and thermal
          evaporation, the clouds survive only a few tens of Myr, but
          during this time they act as an additional cooling agent and
          the internal energy of cloudy models is typically reduced by
          20 -- 40\% in comparison with models without clouds.  The
          clouds delay the development of large-scale outflows,
          therefore helping to retain a larger amount of gas inside
          the galaxy.  However, especially in models with continuous
          creation of infalling clouds, their bullet effect can pierce
          the expanding supershell and create holes through which the
          superbubble can vent freshly produced metals.  Moreover,
          assuming a pristine chemical composition for the clouds,
          their interaction with the superbubble dilutes the gas,
          reducing the metallicity (by up to $\sim$ 0.4 dex) with
          respect to the one attained by diffuse models.  }

\maketitle

\section{Introduction}

Intense episodes of star formation (SF) in dwarf galaxies (DGs) are
associated to the development of large-scale outflows. These have
enormous multiple effects on the evolution of DGs: the hot gas carries
off a large amount of freshly produced metals, thermalizes the
interstellar medium (ISM), reduces the gas content, ceases the SF, and
further more.  The study of this phenomenon through numerical
simulations has therefore been in the focus of several authors in the
past.  The overall picture is that the occurrence of outflows is
initially driven by the thermal pressure of a hot, high pressurized
gas and is favored by a flat distribution of the ISM, which allows an
easy vertical transport of material, whereas horizontal transport is
very limited.  Therefore, outflows are able to eject a large fraction
of freshly produced metals but only a very limited fraction of ISM
(D'Ercole \& Brighenti
\cite{db99}; Recchi, Matteucci \& D'Ercole \cite{rmd}).

Most of these studies, however, consider a single-phase ISM, although
attempts to perform multiphase hydrodynamical simulations have been
made in the past, particularly using the so-called {\it
chemodynamical} approach (Samland, Hensler \& Theis \cite{sht97};
Hensler \cite{hen03}; Hensler, Theis \& Gallagher \cite{hen04}).  In
two previous papers, we have studied the dynamical and chemical
evolution of model galaxies similar to IZw18 (Recchi et
al. \cite{rec04}, hereafter Paper I) and NGC1569 (Recchi et
al. \cite{rec06}).  In this work we simulate the same two galaxies,
but we increase arbitrarily the gas density of some specific regions
of the computational grid, in order to create a ``cloudy'' phase, and
we address the question how and to which extent this new component
alters the former results.

It is particularly interesting to compare the chemical evolution of
models with and without clouds.  In fact, clouds are expected to
dilute the hot metal-rich gas through evaporation, allowing for a
reduction of the metallicity without altering the abundance ratios
(K\"oppen \& Hensler \cite{kh05}).  On the other hand, the clouds can
act as a cap and hamper the development of a galactic wind by means of
their drag and, if they are dispersed, due to Kelvin-Helmholtz
instability and evaporation of mass load to the hot outflow.  As a
result these should lead to an {\it increase} of the metallicity of
the galactic ISM, since, as we have seen, galactic winds otherwise
carry metal-enriched gas away from the galaxy.  In this case, the
abundance ratios are affected if the ejection efficiencies depend on
the different chemical species.

\section{The model}

The simulations are performed by means of a 2-D chemodynamical code
described in Paper I and references therein.  It assumes cylindrical
symmetry and follows in detail the release of mass, individual
elements and energy from SNeIa, SNeII and stellar winds.  As mentioned
in the introduction, the initial set-up of the models is aimed at
reproducing the main characteristics of two well-studied gas-rich
dwarf galaxies, namely IZw18 and NGC1569.

The cloudy phase is introduced by randomly choosing grid cells and
increasing artificially the density therein and in the neighboring
cells in order to reproduce a cloud density profile similar to the
observed ones (namely $\rho_{\rm cl} \propto R_{\rm cl}^{-1.7}$, where
$R_{\rm cl}$ is the distance from the center of the cloud; de Heij,
Braun \& Burton \cite{deh02}).  We assume that the chemical
composition of the clouds is pristine (i.e. without metals).  The
clouds can be either added at the beginning of the simulation ({\it
static cloudy} models) or continuously created during the evolution of
the model.  In the latter case, we will give also an infall velocity
of 10 km s$^{-1}$ to the clouds ({\it infall cloudy} models).  We will
also consider as comparison models obtained without a cloudy phase
({\it diffuse} models).

Since saturated heat conduction is included (\cite{cmk77}) in the
static cloudy model the clouds can only act until they vanish due to
evaporation, while in the infall cloudy model clouds are reborn on
timescales of a few Myr.

\section{Results}

\subsection{Dynamical evolution}
\label{sec:dynres}

\begin{figure}
\includegraphics[width=9cm]{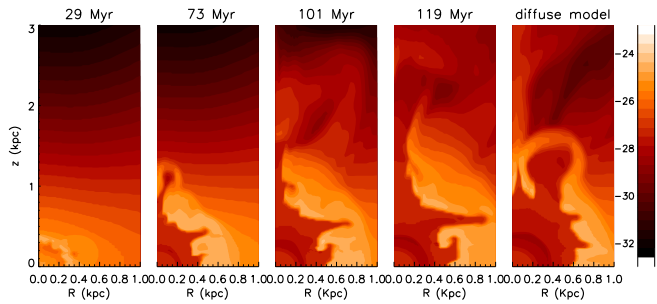}
\caption{Density contours of the gas for the static cloudy model at 
4 evolutionary times (labelled, in Myr, on the top of each panel). The
logarithmic density scale (in g cm$^{-3}$) is on the right-hand
strip. For reference, also the density contours of the correspondent
diffuse model at $t$ $\sim$ 120 Myr is displayed (rightmost panel).}
\label{static}
\end{figure}

The snapshots of the evolution of a prototypical static cloudy model
are shown in Fig. \ref{static} (first 4 panels) together with the
density distribution on the analogous diffuse model at $t$ $\sim$ 120
Myr (rightmost panel).  This model is aimed at reproducing IZw18 and
has a SF history characterized by a mild, long-lasting episode of SF
and a recent burst (see Paper I; Recchi \& Hensler \cite{rh07} for
details).  The clouds are soon engulfed in the hot cavity and
evaporate on a short timescale of a few tens of Myrs. The presence of
the clouds is however sufficient to create large shears and
eddies. Due to these structures and due to the evaporation of the
clouds within the superbubble, radiative losses in this model are
significantly larger than in the equivalent diffuse model (by $\sim$
20 \%).  

At the beginning therefore the evolution of the superbubble is
slightly slower than in diffuse models.  In particular, a galactic
outflow starts $\sim$ 75 Myr after the beginning of the SF; $\sim$
5--10 Myr later than the analogous diffuse model.  However, the
presence of clouds strongly distorts the shape of the supershell and
the highly pressurized gas inside the cavity can more easily find
regions of lower pressure, from which it is easy to pierce the shell
and break out.  This creates the tongues visible in Fig.~\ref{static}.
From these tongues freshly produced metals can leave the galaxy.
Therefore, in spite of the (slightly) slower development of the
galactic wind, the static cloudy model maintains a large ejection
efficiency of newly synthesized metals.

\begin{figure}
\includegraphics[width=9cm]{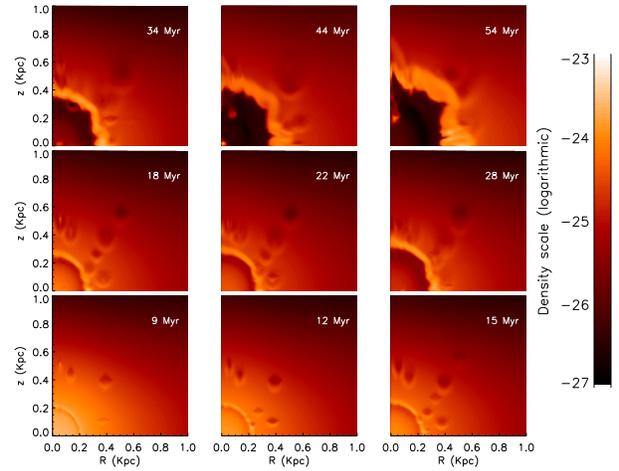}
\caption{Density contours for the infall cloudy model at 9 evolutionary 
times (labeled, in Myr, at the top right corner of each panel). The
logarithmic density scale (in g cm$^{-3}$) is on the right-hand strip.}
\label{infall}
\end{figure}

A similar behavior is seen in the infall cloudy model
(Fig. \ref{infall}).  This model has the same setup as the static
cloudy model seen above, the only difference being the treatment of
clouds.  Bow shocks created by the clouds while falling towards the
disk are evident in this figure.  The interaction between clouds and
ISM produces also Kelvin-Helmholtz instabilities and evaporation,
therefore, also in this model, clouds do not survive more than a few
tens of Myr.  The radiative losses are more significant than in the
static cloudy model because continuously formed clouds deliver more
mass to the outflow.  At the end of the simulation the thermal budget
of the system is reduced by $\sim$ 35\% compared to the analogous
diffuse model.  Consequently, the first weak signs of a galactic
outflow appear only $\sim$ 90 Myr after the beginning of the SF;
$\sim$ 20--25 Myr later than in the diffuse model.  However,
analogously to the case of the static cloudy model, the ram pressure
of the clouds pierces the supershell and creates holes through which
hot cavity gas (and freshly produced metals) can leak out.

\subsection{Chemical evolution}

\begin{figure}
\includegraphics[width=9cm]{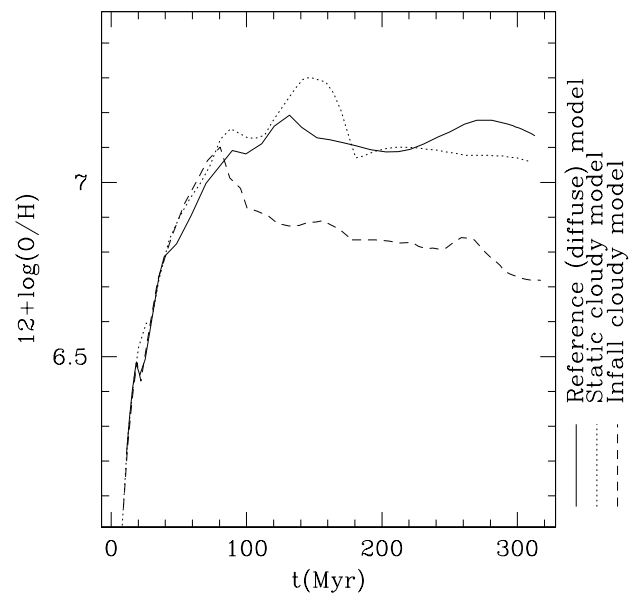}
\caption{Evolution of 12 + log (O/H) for the static cloudy model 
(dotted line), the infall cloudy model (dashed line) and the reference 
diffuse model (solid line).}
\label{cloudo}
\end{figure}

The evolution of the mean oxygen abundance in the ISM with time for
the static cloudy model and the infall cloudy model is shown in
Fig. \ref{cloudo}, in comparison with the evolution of the reference
diffuse model.  The interaction of the clouds with the expanding
supershell has a {\it non-steady} behavior, resulting in moments in
which the attained average metallicity of the galaxy is larger than in
the diffuse model and moments in which it is lower.  As we have
explained in the Introduction, this depends essentially on the
competition between the dilution effect (due to the evaporation of the
metal-free clouds, which tend to reduce the metallicity) and the {\it
drag} effect (the delayed production of galactic outflows, which tends
to increase the metallicity).

At the end of the simulation, however, the metallicity of the cloudy
models is lower than the one of the diffuse model (by $\sim$ 0.1 dex
for the static cloudy model; $\sim$ 0.4 dex for the infall cloudy
one).  This shows that over the long term, the dilution effect
prevails over the restraint effect of the metal-enriched gas and the
clouds have on average the effect of reducing the metallicity.  This
is caused by two facts: on the one hand, (as explained in
Sect. \ref{sec:dynres}) the interaction clouds-supershell distorts the
supershell and produces fingers and tongues through which the metals
present in the cavity can leave the galaxy, therefore the ejection
efficiency of metals is not significantly reduced (in some case it is
even larger) compared to the diffuse model.  The second effect of metal
dilution due to cloud evaporation becomes discernible by the tendency
that in the infall cloudy model the oxygen abundance is continuously
reduced with time after a maximum.

Models reproducing NGC1569 show the same kind of behavior explained
above, therefore we do not describe them in this contribution.  A
comprehensive description of these models (and of others in which the
IMF and the nucleosynthetic yields are varied) can be found in Recchi
et al. (\cite{rec06}).

\section{Conclusions}

Although models of DG winds show strikingly that the low effective
yield causes the low metallicity in dIrrs (Garnett \cite{gar02}),
nevertheless, it is obvious that a vertically extended gas disk and/or
an enveloping gas reservoir is able to hamper the outflow of hot
metal-rich supernova gas.  The presented results demonstrate that the
effects of existing embedded gas clouds and their evaporation on the
effective yield is not straightforward predictable but depends on the
porosity of the gas envelop and its reservoir, i.e. on the delivery of
clouds, and must be studied in detail.

Due to dynamical processes and thermal evaporation, the clouds survive
only a few tens of Myr. Due to the additional cooling agent, the
energy of cloudy models is reduced by 20-40\% compared with diffuse
ones. The clouds delay the development of large-scale outflows,
helping in retaining a larger amount of gas inside the
galaxy. However, their bullet effect pierces the expanding supershell
and creates holes through which the superbubble can vent out the
freshly produced metals. The resulting final metallicity is generally
smaller (by $\sim$ 0.4 dex at most, depending on the model) than the
one attained by diffuse models.

Nevertheless, models with self-regulated SF and self-consistent
treatment of the ISM, in particular with respect to cloudy and hot
phases, are requested to evaluate the chemical abundances reliably.

\acknowledgements
The organizers of the Splinter Meeting are acknowledged for putting
together a very enjoyable and interesting meeting.  S.R. acknowledges
financial support from the Deutsche Forschungsgemeinschaft (DFG) under
grant HE 1487/28-1.

\end{document}